# A novel theorem on motion stability

A.R. Tavakolpour-Saleh[*]

Department of Mechanical and Aerospace Engineering, Shiraz University of Technology, Shiraz, Iran

**Abstract**

Determination of stability and instability of singular points in nonlinear dynamical systems is an important issue that has attracted considerable attention in different fields of engineering and science. So far, different well-defined theories have been presented to study the stability of singular points among which the Lyapunov theory is well-known. However, the instability problem of singular points has been neglected to some extent in spite of its application in oscillator design. Besides, it is often difficult to achieve a proper Lyapunov function for a given complex system. This work presents a novel theorem based on defining two distinct functionals and some straightforward criteria to study motion stability that significantly facilitate the determination of equilibrium status at singular points without the requirement to analytical solution. Indeed, this method is applicable to both stability and instability problems of linear and nonlinear dynamical systems. In addition, the presented theorem is further extended to achieve a new linearization approach of dynamical systems based on averaging technique, which is superior to the Jacobian approach. Lastly, the proposed linearization method is generalized to study the stability/instability of higher-order liner/nonlinear systems. The obtained results clearly show the effectiveness of the proposed theorem to assess the motion stability/instability.

## 1. Introduction

The problem of equilibrium status plays an important role in various fields of physical science such as nonlinear dynamics and chaos [1, 2], complex systems [3], control theories [4, 5], dynamical and biological systems [6]. So far, different stability criteria have been presented to assess the equilibrium status of the singular points such as circle criterion [3], Jury stability criterion [7], Routh–Hurwitz criterion [8], Nyquist criterion [8], Liénard–Chipart criterion [9], Vakhitov–Kolokolov stability criterion [10], Barkhausen stability criterion [11] etc. However, the first attempt to present a global approach to analyze the stability of nonlinear systems was made by Lyapunov in 1892 [2, 12]. Chetaev was a Russian mechanician who first realized the significance of the Lyapunov theory [13].

---

[*] Corresponding author. Tell.: +98 9173147706. Fax: +987137264102.
*E-mail address:* tavakolpour@sutech.ac.ir; alitavakolpur@yahoo.com

The Lyapunov theory contains two approaches to show stability of an equilibrium state in a dynamical system, namely indirect and direct methods [4, 5]. The indirect method of Lyapunov states that the stability of a nonlinear system around a singular point is identical to that of its linearized approximation. On the other hand, the direct method that is the main focus of the Lyapunov stability theory expresses if the total mechanical energy of a system decreases continuously and ultimately, reaches zero or a certain value, then, the system will approach the singular point or stay near it which means that the system is stable [4]. This latest technique relies on first defining a measure of total energy in a dynamical system that is called the Lyapunov function and second, the rate of change of the so-called Lyapunov function is studied to ascertain the system stability. It is worth noting that such a time derivative of the energy function is indeed the instantaneous power of a dynamical system. It means that the standard Lyapunov theory considers the sign of instantaneous power of the system to determine stability. The main advantage of this technique can be attributed to its generality. Conversely, the main challenge of this powerful tool lies in the fact that it is sometimes troublesome to reach a Lyapunov function for a given system [4]. Almost, the majority of theories presented on dynamical systems have been focused on the stability criteria. Nevertheless, in practice, there are some systems, in which instability criteria is required to justify their operation (e.g. free piston engines [14, 15] and oscillators [16-18]). However, just a few researches can be found on the application of Lyapunov theory for investigating the instability of some limited dynamical systems implying that the standard Lyapunov theory is not so effective for instability problems [19-20]. Tavakolpour-Saleh and Zare [21] proposed an extended version of the Lyapunov theory. They then showed the effectiveness of the proposed theorem compared to the well-defined Lyapunov theorem to prove the instability of equilibrium point in a complex coupled oscillator [21, 22]. Although, this latest theorem simplified the original Lyapunov criteria so as to achieve more alternative functions, it still relied on finding a suitable scalar function to prove stability/instability of a fixed point.

In this work, in order to overcome the drawbacks of the conventional Lyapunov theory, a more straightforward method justifying both instability and stability of motion near singular points of dynamical systems is presented. The main idea is to introduce two distinct functionals associated with some distinct criteria resulting in a new theorem that is a powerful alternative for the so-called Lyapunov theorem. Besides, a new linearization scheme based on averaging technique is introduced and then, generalized for investigating stability/instability of linear/nonlinear systems (at the vicinity of the origin). Finally, the validity as well as the generality of the proposed theory is verified.

## 2. Physical and mathematical backgrounds

The main idea of this work came into being by deeply realizing the startup process of the combustion engines and mechanical oscillators. Accordingly, in order to realize the paper idea it is worthwhile to look at a very old car equipped with a starting handle. Let's see how driver can start such an old car's engine.

First, the driver has to rotate the engine's crankshaft via the attached starting handle at least one perfect revolution (i.e. $2\pi$ rad) with an arbitrary low speed (e.g. 1 rad/s). This initial rotational motion of the crankshaft results in the reciprocating motion of a piston that can be defined by a periodic function as $x=\cos(t)$ where $x$ is the piston's coordinate. In this process, the driver exerts an initial periodic motion pattern to the engine system as an external excitation although it is not a realistic system trajectory. However, if the average value of the generated instantaneous power inside the engine over one thermodynamic cycle (due to the fuel ignition in the power stroke) exceeds the average frictional power and the average power required for the compression stroke, the engine starts to run continuously with a higher frequency (meaning that the system equilibrium point is unstable). In contrast, if the average generated power (over one cycle) does not reach the summation of the average frictional power and the average power required for compression stroke (over one cycle), the engine will be shut down (meaning that the system equilibrium point is stable).

As another example to clarify the paper idea, consider reciprocating oscillators such as free piston Stirling engines [14, 15] or Van der Pol oscillator [16, 17]. One can excite such oscillators at the vicinity of its equilibrium point by exerting an initial reciprocating motion (or initial orbit) of small amplitude (i.e. $\varepsilon \to 0^+$) and arbitrary angular frequency (e.g. 1 rad/s which results in a period of $2\pi$ rad) such as $x = \varepsilon\cos(t)$ (and consequently, $\dot{x} = -\varepsilon\sin(t)$). Therefore, if the average generated power in one working cycle ($2\pi$ rad) is more than the average dissipated and consumed powers, the oscillator equilibrium point (origin) will be unstable. These latest examples presents a new idea of exerting an initial excitation orbit (i.e. $x = \varepsilon\cos(t)$ and $\dot{x} = -\varepsilon\sin(t)$) enclosing the origin with small radios ($\varepsilon$) to excite a dynamical system rather than the conventional initial conditions ($x = x_0$ and $\dot{x} = \dot{x}_0$).

**Remark.** It is important to note that the considered excitation orbit enclosing the origin is not a realistic system trajectory; whereas, it is a sort of external excitation and thus, can have any forms e.g. $x = \varepsilon\cos(t)$ and $\dot{x} = -\varepsilon\sin(t)$. In other words, a dynamical system can be externally excited by an initial orbit enclosing the origin with a sufficiently small radius and then, the stability or instability of the origin can be determined by looking at the change of system energy which will be discussed in the next section.

As mentioned repeatedly in this section, the average of instantaneous power plays an important role in the determination of stability of a dynamical system around the origin. Besides, the use of average power instead of the instantaneous power is a conventional technique in different fields of engineering such as acoustic [21]. Consequently, due to the importance of averaging operation in the presented work, it is essential to define the averaging operator from mathematical point of view. In general, the average value of a periodic function (denoted by $g(t)$), over one working cycle about point $t$ can be expressed as [1, 21, 22]

$$\langle g \rangle_t = \frac{1}{2\pi} \int_{t-\pi}^{t+\pi} g(s)\, ds \tag{1}$$

where $\langle.\rangle$ is the averaging operator and *s* is dummy integral variable. The process described by Eq. (1) is sometimes called running averaging as *t* is explicitly considered in the integral bands. However, at a specific time, the average value of a periodic function can be stated as

$$\langle g \rangle = \frac{1}{2\pi}\oint_0^{2\pi} g(s)\, ds \tag{2}$$

It is interesting to note that although the averaging process is carried out over time, the cyclic integrating process defined by Eq. (2) will be independent of time. In addition, the average value of function *g(t)* over one cycle (i.e. $2\pi$ rad) is independent of frequency ($\langle \cos(w_1 t)\rangle_{2\pi} = \langle \cos(w_2 t)\rangle_{2\pi} = 0$). Similarly, it is also independent of the function period ($\langle \cos(\frac{2\pi}{T_1} t)\rangle_{2\pi} = \langle \cos(\frac{2\pi}{T_2} t)\rangle_{2\pi} = 0$). Indeed, the background presented in this section is significant in understanding the paper idea discussed in the following subsection.

## 3. Theorem

For a major class of nonlinear/linear dynamical system possessing the motion equation as $\ddot{x} = f(x, \dot{x})$ with an equilibrium state at the origin of the phase plane, two functionals denoted by $\mathcal{T}_1$ and $\mathcal{T}_2$ are defined as

$$\mathcal{T}_1(f(x,\dot{x})) = \langle 2\dot{x} f(x,\dot{x})\rangle_{2\pi} \quad \text{in which } x = \varepsilon\cos(t) \text{ and } \dot{x} = -\varepsilon\sin(t) \tag{3}$$

$$\mathcal{T}_2(f(x,\dot{x})) = \langle 2x f(x,\dot{x})\rangle_{2\pi} \quad \text{in which } x = \varepsilon\cos(t) \text{ and } \dot{x} = -\varepsilon\sin(t) \tag{4}$$

where $\varepsilon$ is the radius of a small excitation orbit enclosing the origin. Then, the general status of equilibrium point (origin) can be determined based on the criteria given in Table 1.

**Table 1** Presented criteria for determination of the equilibrium status

| Criterion No. | Criteria | Equilibrium status |
|---|---|---|
| (I) | $\mathcal{T}_1 > 0$ | Unstable |
| (II) | $\mathcal{T}_1 < 0$ and $\mathcal{T}_2 \geq 0$ | Unstable |
| (III) | $\mathcal{T}_1 < 0$ and $\mathcal{T}_2 < 0$ | Asymptotically Stable |
| (IV) | $\mathcal{T}_1 = 0$ and $\mathcal{T}_2 < 0$ | Marginally Stable |
| (V) | $\mathcal{T}_1 = 0$ and $\mathcal{T}_2 \geq 0$ | Unstable |

**Proof.** Since all $2^{ed}$-order nonlinear dynamic systems behave linearly at the vicinity of the origin ($\varepsilon$ is small), let's consider a linear differential equation with real eigenvalues $\lambda_1$ and $\lambda_2$ as

$$\ddot{x} - f(x,\dot{x}) = 0 \tag{5}$$

where

$$f(x, \dot{x}) = (\lambda_1 + \lambda_2)\dot{x} - \lambda_1\lambda_2 x \tag{6}$$

Clearly, the origin is the equilibrium state of Eq. (5). Multiplying both sides of Eq. (5) by $\dot{x}$ yields

$$\dot{x}\ddot{x} - \dot{x}f(x,\dot{x}) = 0 \tag{7}$$

It is obvious that the term $\dot{x}\ddot{x}$ in Eq. (7) is the instantaneous power of the mass term in the dynamic system, which is indeed the time derivative of the kinetic energy ($\dot{x}\ddot{x} = \frac{d}{dt}(\frac{1}{2}\dot{x}^2)$). However, according to the proposed idea, the average power over a very small excitation orbit enclosing the origin ($x = \varepsilon\cos(t)$ and $\dot{x} = -\varepsilon\sin(t)$) is used here instead of the instantaneous power considered in the Lyapunov theory ($\dot{V}$). Consequently, applying averaging operator over one cycle ($2\pi$ rad) to both sides of Eq. (7) leads to

$$\langle \dot{x}\ddot{x} \rangle_{2\pi} = \langle \dot{x}f(x,\dot{x}) \rangle_{2\pi} \tag{8}$$

Substituting Eq. (6) into Eq. (8) and considering a small excitation orbit enclosing the origin as $x = \varepsilon\cos(t)$ and $\dot{x} = -\varepsilon\sin(t)$, the right-hand side of Eq. (8) can be expressed as

$$\langle \dot{x}f(x,\dot{x}) \rangle_{2\pi} = \langle \dot{x}((\lambda_1 + \lambda_2)\dot{x} - \lambda_1\lambda_2 x) \rangle_{2\pi} = (\lambda_1 + \lambda_2)\varepsilon^2 \langle \sin^2(t) \rangle_{2\pi} + \lambda_1\lambda_2\varepsilon^2 \langle \sin(t)\cos(t) \rangle_{2\pi} \tag{9}$$

Subsequently, Eq. (9) can be simplified based on the average values of trigonometric functions (i.e. $\langle \cos(t)\sin(t) \rangle_{2\pi} = 0$ and $\langle \cos^2(t) \rangle_{2\pi} = \langle \sin^2(t) \rangle_{2\pi} = 1/2$ ) as

$$\langle \dot{x}\ddot{x} \rangle_{2\pi} = \langle \dot{x}f(x,\dot{x}) \rangle_{2\pi} = \frac{1}{2}(\lambda_1 + \lambda_2)\varepsilon^2 + 0 \tag{10}$$

And thus,

$$\langle 2\dot{x}\ddot{x} \rangle_{2\pi} = \langle 2\dot{x}f(x,\dot{x}) \rangle_{2\pi} = (\lambda_1 + \lambda_2)\varepsilon^2 \tag{11}$$

Eq. (11) states that the first proposed functional $\mathcal{T}_1 = \langle 2\dot{x}f(x,\dot{x}) \rangle_{2\pi}$ returns the summation of eigenvalues of the dynamical system. It is interesting to note that the system eigenvalues directly appear in $\mathcal{T}_1$ and thus, evaluating $\mathcal{T}_1$ over the small excitation orbit enclosing the origin serves to provide important information on the instability (or stability) of the origin in a dynamic system. Consequently, once $\mathcal{T}_1 > 0$, different cases for eigenvalues can be expected as given in Table 2. Accordingly, if $\mathcal{T}_1$ is found to be positive over the small excitation orbit enclosing the origin, one of the three mentioned cases (see Table 2) may occur in which the system origin is unstable. Consequently, it is proved that the sufficient condition for instability of the origin is $\mathcal{T}_1 > 0$. Now, it is significant to study the equilibrium status once $\mathcal{T}_1 < 0$. Table 3 describes the possible cases in which $\mathcal{T}_1 < 0$. Accordingly, in the first two cases the system origin will be stable while for the third case the system origin is unstable. As a result, another criterion is required to determine stability or instability of the origin once $\mathcal{T}_1 < 0$.

**Table 2** different situations of eigenvalues that results in $\mathcal{T}_1 > 0$

| Case No. | Conditions | Sign of $\mathcal{T}_1$ for a very small orbit enclosing the origin | Status of the origin |
|---|---|---|---|
| Case I | $\text{Re}(\lambda_1) > 0, \text{Re}(\lambda_2) > 0, \text{Re}(\lambda_1) \neq \text{Re}(\lambda_2)$ | $\lambda_1 + \lambda_2 > 0 \rightarrow \mathcal{T}_1 > 0$ | Unstable |
| Case II | $\text{Re}(\lambda_1) > 0, \text{Re}(\lambda_2) > 0, \text{Re}(\lambda_1) = \text{Re}(\lambda_2)$ | $2\lambda_1 > 0 \rightarrow \mathcal{T}_1 > 0$ | Unstable |
| Case III | $\text{Re}(\lambda_1) < 0, \text{Re}(\lambda_2) > 0, \text{Re}(\lambda_2) > -\text{Re}(\lambda_1)$ | $\lambda_1 + \lambda_2 > 0 \rightarrow \mathcal{T}_1 > 0$ | Unstable |

In order to achieve another criterion, both sides of Eq. (5) is multiplied by *x* which results in

$$x\ddot{x} - xf(x, \dot{x}) = 0 \tag{12}$$

Applying averaging operator over one cycle (2π rad) to both sides of Eq. (12) gives

$$\langle x\ddot{x} \rangle_{2\pi} = \langle xf(x, \dot{x}) \rangle_{2\pi} \tag{13}$$

Substituting Eq. (6) into Eq. (13) and considering a small orbit enclosing the origin as $x = \varepsilon\cos(t)$ and $\dot{x} = -\varepsilon\sin(t)$, the right-hand side of Eq. (13) can be stated as

$$\langle xf(x, \dot{x}) \rangle_{2\pi} = \langle x((\lambda_1 + \lambda_2)\dot{x} - \lambda_1\lambda_2 x) \rangle_{2\pi} = (\lambda_1 + \lambda_2)\varepsilon^2 \langle \cos(t)\sin(t) \rangle_{2\pi} - \lambda_1\lambda_2\varepsilon^2 \langle \cos^2(t) \rangle_{2\pi} \tag{14}$$

Subsequently, Eq. (14) can be simplified based on the average values of trigonometric functions (i.e. $\langle \cos(t)\sin(t) \rangle_{2\pi} = 0$ and $\langle \cos^2(t) \rangle_{2\pi} = \langle \sin^2(t) \rangle_{2\pi} = 1/2$ ) as

$$\langle xf(x, \dot{x}) \rangle_{2\pi} = 0 - \frac{1}{2}(\lambda_1\lambda_2)\varepsilon^2 \tag{15}$$

And thus,

$$\langle 2x\ddot{x} \rangle_{2\pi} = \langle 2xf(x, \dot{x}) \rangle_{2\pi} = -(\lambda_1\lambda_2)\varepsilon^2 \tag{16}$$

As seen in Eq. (16), the second criterion $\mathcal{T}_2 = \langle 2xf(x,\dot{x})\rangle_{2\pi}$ of the proposed theorem contains the product of eigenvalues (real part of eigenvalues). Looking at Table 3, it is interesting to note that in the third case where the origin is unstable, $\mathcal{T}_2 > 0$ while in the first two cases where the systems are stable, $\mathcal{T}_2 < 0$. As a consequence, once $\mathcal{T}_1 < 0$ another criterion on $\mathcal{T}_2$ must be checked to assess the stability or instability of the origin. Although, in the theorem proof the eigenvalues were considered real, however, the consideration of complex conjugate eigenvalues leads to the second cases of Tables 2 and 3 if real parts of the complex eigenvalues are considered in the theorem proof.

**Remark.** It is worth noting that although the 2$^{\text{ed}}$ order linear systems were considered in the proof of the theorem, a same implication can be inferred for the 2$^{\text{ed}}$ order nonlinear systems because all nonlinear systems behave linearly at the vicinity of their equilibrium points.

**Table 3** different situations of eigenvalues once $\mathcal{T}_1 < 0$

| Case No. | Conditions | Sign of $\mathcal{T}_1$ and $\mathcal{T}_2$ for a small orbit enclosing the origin | Status of the origin |
|---|---|---|---|
| Case I | $\text{Re}(\lambda_1) < 0, \text{Re}(\lambda_2) < 0, \text{Re}(\lambda_1) \neq \text{Re}(\lambda_2)$ | $\lambda_1 + \lambda_2 < 0 \rightarrow \mathcal{T}_1 < 0$ | Asymptotically Stable |
| | | $-\lambda_1\lambda_2 < 0 \rightarrow \mathcal{T}_2 < 0$ | |
| Case II | $\text{Re}(\lambda_1) < 0, \text{Re}(\lambda_2) < 0, \text{Re}(\lambda_1) = \text{Re}(\lambda_2)$ | $2\lambda_1 < 0 \rightarrow \mathcal{T}_1 < 0$ | Asymptotically Stable |
| | | $-\lambda_1^2 < 0 \rightarrow \mathcal{T}_2 < 0$ | |
| Case III | $\text{Re}(\lambda_1) < 0, \text{Re}(\lambda_2) > 0, \text{Re}(\lambda_2) < -\text{Re}(\lambda_1)$ | $\lambda_1 + \lambda_2 < 0 \rightarrow \mathcal{T}_1 < 0$ | Unstable |
| | | $-\lambda_1\lambda_2 > 0 \rightarrow \mathcal{T}_2 > 0$ | |

## 4. Generalization of the presented theorem for 2$^{\text{ed}}$-order systems

For the general form of 2$^{\text{ed}}$-order nonlinear/linear dynamical systems in the state-space such as

$$\dot{x}_1 = f_1(x_1, x_2) \qquad (17)$$
$$\dot{x}_2 = f_2(x_1, x_2) \qquad (18)$$

The equilibrium status of the origin in the phase plane can be determined based on two functionals denoted by $\mathcal{T}_1$ and $\mathcal{T}_2$ as

$$\mathcal{T}_1(f_1, f_2) = 2(\langle x_1 f_1\rangle_{2\pi} + \langle x_2 f_2\rangle_{2\pi}) \qquad (19)$$
$$\mathcal{T}_2(f_1, f_2) = 2(\langle x_2 f_1\rangle_{2\pi} \times \langle x_1 f_2\rangle_{2\pi} - \langle x_1 f_1\rangle_{2\pi} \times \langle x_2 f_2\rangle_{2\pi}) \qquad (20)$$

where $x_1 = \varepsilon\cos(t)$ and $x_2 = -\varepsilon\sin(t)$ and $\varepsilon$ is the radius of a very small excitation orbit enclosing the origin. Then, the criteria given in Table 1 can be used to determine the equilibrium status of the origin.

**Proof.** As all nonlinear systems can be considered as linear at the vicinity of an equilibrium point at the origin, let's consider a general form of linear dynamic system as $\dot{\mathbf{x}} = \mathbf{A}\mathbf{x}$ with the equilibrium state at the origin (i.e. $\mathbf{x_e}$=(0,0)) such that

$$\begin{bmatrix} \dot{x}_1 \\ \dot{x}_2 \end{bmatrix} = \begin{bmatrix} a & b \\ c & d \end{bmatrix} \begin{bmatrix} x_1 \\ x_2 \end{bmatrix} \rightarrow \begin{cases} \dot{x}_1 = ax_1 + bx_2 = f_1 \\ \dot{x}_2 = cx_1 + dx_2 = f_2 \end{cases} \tag{21}$$

where $\mathbf{x} = [x_1\ x_2]^T = [x\ \dot{x}]^T$ is the state vector and $\mathbf{A}= \begin{bmatrix} a & b \\ c & d \end{bmatrix}$ is the state matrix. From the knowledge of linear algebra, the eigenvalues of the system can be acquired using the following equation [1]

$$\lambda^2 - \tau\lambda + \Delta = 0 \tag{22}$$

where

$$\tau = trace(A) = a + d = \lambda_1 + \lambda_2 \tag{23}$$
$$\Delta = \det(A) = ad - bc = \lambda_1\lambda_2 \tag{24}$$

Now, based on the idea of exciting the system with a small excitation orbit enclosing the origin ($x_1 = \varepsilon\cos(t)$ and $x_2 = -\varepsilon\sin(t)$), let's define the following terms as

$$\langle x_1\dot{x}_1\rangle_{2\pi} = \langle ax_1^2 + bx_1x_2\rangle_{2\pi} = \langle a\varepsilon^2\cos(t)^2\rangle_{2\pi} - \langle b\varepsilon^2\cos(t)\sin(t)\rangle_{2\pi} = \frac{\varepsilon^2}{2}a - 0 \tag{25}$$

$$\langle x_2\dot{x}_1\rangle_{2\pi} = \langle ax_2x_1 + bx_2^2\rangle_{2\pi} = -\langle a\varepsilon^2\sin(t)\cos(t)\rangle_{2\pi} + \langle b\varepsilon^2\cos(t)^2\rangle_{2\pi} = 0 + \frac{\varepsilon^2}{2}b \tag{26}$$

$$\langle x_1\dot{x}_2\rangle_{2\pi} = \langle cx_1^2 + dx_1x_2\rangle_{2\pi} = \langle c\varepsilon^2\cos(t)^2\rangle_{2\pi} - \langle d\varepsilon^2\cos(t)\sin(t)\rangle_{2\pi} = \frac{\varepsilon^2}{2}c - 0 \tag{27}$$

$$\langle x_2\dot{x}_2\rangle_{2\pi} = \langle cx_2x_1 + dx_2^2\rangle_{2\pi} = -\langle c\varepsilon^2\sin(t)\cos(t)\rangle_{2\pi} + \langle d\varepsilon^2\cos(t)^2\rangle_{2\pi} = 0 + \frac{\varepsilon^2}{2}d \tag{28}$$

Substituting Eqs. (25) and (28) into Eq. (23) yields

$$(\lambda_1 + \lambda_2)\varepsilon^2 = (a+d)\varepsilon^2 = 2(\langle x_1\dot{x}_1\rangle_{2\pi} + \langle x_2\dot{x}_2\rangle_{2\pi}) \tag{29}$$

And substituting Eqs. (25) - (28) into Eq. (24) yields

$$-\lambda_1\lambda_2\varepsilon^2 = -(ad-bc)\varepsilon^2 = 2(\langle x_2\dot{x}_1\rangle_{2\pi} \times \langle x_1\dot{x}_2\rangle_{2\pi} - \langle x_1\dot{x}_1\rangle_{2\pi} \times \langle x_2\dot{x}_2\rangle_{2\pi}) \tag{30}$$

Finally, substituting Eqs. (17) and (18) into Eqs. (29) and (30) leads to the functionals defined by Eqs. (19) and (20). Consequently, it is now possible to use the criteria given in Table 1 to determine stability or instability of the origin.

## 5. Linearization via averaging and its application to reach equilibrium status of higher order systems

Another interesting outcome of the proposed scheme is the linearization of nonlinear systems around the origin through averaging principle. Accordingly, using Eqs. (25) - (28) and considering a small excitation orbit enclosing the origin (i.e. $x_1 = \varepsilon\cos(t)$ and $x_2 = -\varepsilon\sin(t)$ where $\varepsilon$ is small) the state matrix (**A**) of the linearized system ($\dot{\mathbf{x}} = \mathbf{A}\mathbf{x}$) can be expressed as

$$\mathbf{A} = \begin{bmatrix} a & b \\ c & d \end{bmatrix} = \text{Lim}\frac{2}{\varepsilon^2}\begin{bmatrix} \langle x_1 f_1\rangle_{2\pi} & \langle x_2 f_1\rangle_{2\pi} \\ \langle x_1 f_2\rangle_{2\pi} & \langle x_2 f_2\rangle_{2\pi} \end{bmatrix}_{\varepsilon\to 0^+} \tag{31}$$

Similarly, for the general form of the higher-order nonlinear systems in the state-space as

$$\begin{bmatrix} \dot{x}_1 \\ \dot{x}_2 \\ \vdots \\ \dot{x}_n \end{bmatrix} = \begin{bmatrix} f_1(x_1, x_2, \ldots, x_n) \\ f_2(x_1, x_2, \ldots, x_n) \\ \vdots \\ f_n(x_1, x_2, \ldots, x_n) \end{bmatrix} \tag{32}$$

Eq. (31) can be further extended to achieve the state matrix **A** of the linearized system as follows

$$\mathbf{A} = \begin{bmatrix} a_{11} & \cdots & a_{1n} \\ \vdots & \ddots & \vdots \\ a_{n1} & \cdots & a_{nn} \end{bmatrix} = \text{Lim}\frac{2}{\varepsilon^2}\begin{bmatrix} \langle x_1 f_1\rangle_{2\pi} & \cdots & \langle x_n f_1\rangle_{2\pi} \\ \vdots & \ddots & \vdots \\ \langle x_1 f_n\rangle_{2\pi} & \cdots & \langle x_n f_n\rangle_{2\pi} \end{bmatrix}_{\varepsilon\to 0^+} \tag{33}$$

Since the state-space is *n*-dimensional in this case, the terms $\langle x_i f_j\rangle_{2\pi}$ in the state matrix must be calculated considering a small excitation orbit enclosing the origin on the *i-j* plane (meaning that $x_i = \varepsilon\sin(t)$ and $x_j = \varepsilon\cos(t)$ while the rest of state variables are zero). Clearly, the eigenvalues of the state matrix **A** given in Eq. (33) are poles of the linearized dynamical system. Thus,

$$\lambda_i = \text{Eigen}\{\text{Lim}(\frac{2}{\varepsilon^2}\begin{bmatrix} \langle x_1 f_1\rangle_{2\pi} & \cdots & \langle x_n f_1\rangle_{2\pi} \\ \vdots & \ddots & \vdots \\ \langle x_1 f_n\rangle_{2\pi} & \cdots & \langle x_n f_n\rangle_{2\pi} \end{bmatrix}_{\varepsilon\to 0^+})\} \tag{34}$$

Consequently, if all eigenvalues ($\lambda_i$) had negative real parts the system would be stable; otherwise, the linearized system would be unstable (or marginally stable).

**Remark.** At this moment, an important question may arise whether Eq. (33) is identical to the Jacobian matrix or not. Obviously, it is different from the Jacobian matrix; however, once $\varepsilon \to 0^+$, Eq. (33) will approach the Jacobian matrix. Therefore, an innovative linearization approach based on averaging process is proposed instead of the well-defined Jacobian matrix. Next, another question may arise about the effectiveness of the proposed averaging-based linearization approach compared to the Jacobian technique. As clearly mentioned in different text books on nonlinear systems and control theory [4, 5], once the linearized system obtained from the Jacobian method was marginally stable no conclusion could be drawn about the stability, instability, or marginal stability of the original nonlinear systems. However, the proposed averaging-based linearization approach will eliminate this drawback. Indeed, Eq. (33) can still be evaluated for orbits in which $\varepsilon > 0$ meaning that once $\varepsilon$ is increased gradually the behavior of the linearized system approaches the original nonlinear system (see Fig. 1). Consequently, the position of the eigenvalues in the s-plane can vary as a function of the orbit radius ($\varepsilon$) as

$$\lambda_i(\varepsilon^+) = \text{Eigen}(\frac{2}{\varepsilon^2}\begin{bmatrix} \langle x_1 f_1 \rangle_{2\pi} & \cdots & \langle x_n f_1 \rangle_{2\pi} \\ \vdots & \ddots & \vdots \\ \langle x_1 f_n \rangle_{2\pi} & \cdots & \langle x_n f_n \rangle_{2\pi} \end{bmatrix}) \tag{35}$$

Consequently, even if the linearized system is marginally stable (e.g. Fig. 1) one will be able to determine the status of the original nonlinear system around the origin based on the position of system poles corresponding to the increasing value of $\varepsilon$ (i.e. $\varepsilon^+$) using Eq. (35). Finally, the stability of the original nonlinear system can be affirmed if and only if

$$\text{Re}(\lambda_i(\varepsilon^+)) < 0 \tag{36}$$

Overall, it seems that the presented technique combines the indirect and direct Lyapunov methods (i.e. linearization and energy techniques) in order to achieve a unified approach to determine system stability without the requirement to search for a Lyapunov function.

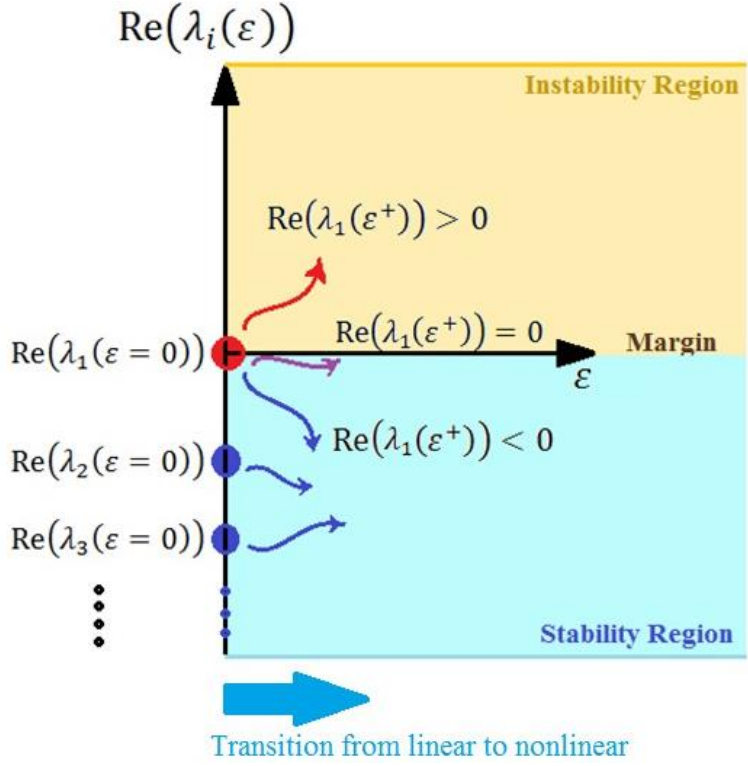

**Fig. 1.** Transition from linear to nonlinear state for increasing value of $\varepsilon$

## 6. Results and discussion

Until now, a new stability theorem based on averaging principle along with a novel linearization technique was introduced and some evidence was given to verify the presented idea. Subsequently, in this section, the proposed theorem is applied to study the motion stability at the vicinity of an equilibrium point. Since the motion equations are usually stated as the second order differential equations based on the Newton's second law, the second order systems are considered in this section.

*6.1. Applying the stability theorem to $2^{ed}$-order linear systems*

In this section, the proposed theorem is applied to different $2^{ed}$-order linear systems to verify its effectiveness. Thus, different case studies are presented in this section to clarify the idea. Table 4 presents a comprehensive research on the application of the theorem for investigating the instability of the origin in different possible $2^{ed}$ order linear systems. One should keep in mind that, the signs of $\mathcal{T}_1$ and $\mathcal{T}_2$ are determined using the average values of the conventional trigonometric terms as

$$\langle \cos(t)\sin(t)\rangle_{2\pi} = 0 \tag{37}$$

$$\langle \cos^2(t)\rangle_{2\pi} = \langle \sin^2(t)\rangle_{2\pi} = \tfrac{1}{2} \tag{38}$$

$$\langle \cos^4(t)\rangle_{2\pi} = \langle \sin^4(t)\rangle_{2\pi} = \tfrac{3}{8} \tag{39}$$

$$\langle \cos^2(t)\sin^2(t)\rangle_{2\pi} = \tfrac{1}{8} \tag{40}$$

As seen in Table 4, for items (1) to (4) the first criterion of the theorem (see Table 1) is fulfilled (i.e. $\mathcal{T}_1 > 0$) which is the sufficient condition to prove instability of the origin (and such instability could be easily realized based on the considered eigenvalues). On the other hand, for items (5) and (6) in which $\mathcal{T}_1 < 0$ the next criterion on $\mathcal{T}_2$ must be checked (see also cases II and V of Table 1) through which the origin instability is affirmed. According to the fifth and sixth cases of Table 4, it is inferred that the criterion II of Table 1 just happens in the saddle points in which $\lambda_1+\lambda_2<0$ and the criterion V just occurs once $\lambda_1=\lambda_2=0$. Table 5 illustrates the possible cases of $2^{ed}$ order linear systems with stable origin. Obviously, the criterion III of Table 1 is fulfilled in the first three examples of Table 5 meaning the origin is stable. In addition to that, the forth case of Table 5 reveals that the criterion IV of Table 1 just happens once the eigenvalues are purely imaginary and thus, the system is marginally stable.

**Remark.** In addition to the determination of stability and instability, the presented theorem can also be conversely used in a practical way to estimate the eigenvalues (real parts of eigenvalues) based on the obtained amounts of $\mathcal{T}_1$ and $\mathcal{T}_2$. Consequently, it is possible to further assess the type of singular points (i.e. saddle, node, focus, center etc.) based on the summarized results given in Table 6. For instance, consider the first row of Table 6. Accordingly, based on the calculated amounts of $\mathcal{T}_1$ and $\mathcal{T}_2$, if one can find two positive real values for parameters $a$ and $b$ justifying the set of equations $\mathcal{T}_1 = -(a+b)\varepsilon^2$ and $\mathcal{T}_2 = -ab\varepsilon^2$, the system origin will be a stable node (nodal sink). The proposed idea can also be applied to nonlinear differential equations of motion as well.

**Table 4** Applying the proposed theorem to different possible unstable linear $2^{ed}$ order systems

| No | System | Eigen values | Sign of $\mathcal{T}_1$ and $\mathcal{T}_2$ (For $x = \varepsilon\cos(t)$ and $\dot{x} = -\varepsilon\sin(t)$) | Origin Status based on the proposed theorem |
|---|---|---|---|---|
| 1 | $\ddot{x} - (a+b)\dot{x} + abx = 0$ | $\lambda_1 = a$<br>$\lambda_2 = b$<br>($0 < a < b$) | $\mathcal{T}_1 = \langle -2\dot{x}f \rangle_{2\pi} = \langle -2\dot{x}(-(a+b)\dot{x} + abx) \rangle_{2\pi}$<br>$= 2(a+b)\langle \dot{x}^2 \rangle_{2\pi} - 2ab\langle \dot{x}x \rangle_{2\pi}$<br>$= 2(a+b)\varepsilon^2\langle \sin^2(t) \rangle_{2\pi} - 2ab\varepsilon^2\langle \sin(t)\cos(t) \rangle_{2\pi}$<br>$= (a+b)\varepsilon^2 > 0$<br>$\{\mathcal{T}_2 = \langle -2xf \rangle_{2\pi} = \langle -2x(-(a+b)\dot{x} + abx) \rangle_{2\pi} = -ab\varepsilon^2 < 0\}$ | Unstable (Criterion I) |
| 2 | $\ddot{x} - 2a\dot{x} + a^2 x = 0$ | $\lambda_1 = \lambda_2 = a$<br>($0 < a$) | $\mathcal{T}_1 = \langle -2\dot{x}f \rangle_{2\pi} = \langle -2\dot{x}(-2a\dot{x} + a^2 x) \rangle_{2\pi}$<br>$= 4a\langle \dot{x}^2 \rangle_{2\pi} - 4a^2\langle x\dot{x} \rangle_{2\pi} = 4a\varepsilon^2\langle \sin^2(t) \rangle_{2\pi} - 0 = 2a\varepsilon^2 > 0$<br>$\{\mathcal{T}_2 = \langle -2xf \rangle_{2\pi} = \langle -2x(-2a\dot{x} + a^2 x) \rangle_{2\pi} = -a^2\varepsilon^2 < 0\}$ | Unstable (Criterion I) |
| 3 | $\ddot{x} - 2a\dot{x} + (a^2 + b^2)x = 0$ | $\lambda_1 = a + bj$<br>$\lambda_2 = a - bj$<br>($0 < a, 0 < b$) | $\mathcal{T}_1 = \langle -2\dot{x}f \rangle_{2\pi} = \langle -2\dot{x}(-2a\dot{x} + (a^2 + b^2)x) \rangle_{2\pi}$<br>$= 4a\langle \dot{x}^2 \rangle_{2\pi} - 2(a^2 + b^2)\langle \dot{x}x \rangle_{2\pi}$<br>$= 4a\varepsilon^2\langle \sin^2(t) \rangle_{2\pi} - 2\varepsilon^2(a^2 + b^2)\langle \sin(t)\cos(t) \rangle_{2\pi} = 2a\varepsilon^2 > 0$<br>$\{\mathcal{T}_2 = \langle -2x(-2a\dot{x} + (a^2 + b^2)x) \rangle_{2\pi} = -(a^2 + b^2)\varepsilon^2 < 0\}$ | Unstable (Criterion I) |
| 4 | $\ddot{x} - (b-a)\dot{x} - abx = 0$ | $\lambda_1 = -a$<br>$\lambda_2 = b$<br>($0 \leq a < b$) | $\mathcal{T}_1 = \langle -2\dot{x}f \rangle_{2\pi} = \langle -2\dot{x}(-(b-a)\dot{x} - abx) \rangle_{2\pi}$<br>$= 2(b-a)\langle \dot{x}^2 \rangle_{2\pi} + 2ab\langle \dot{x}x \rangle_{2\pi}$<br>$= 2(b-a)\varepsilon^2\langle \sin^2(t) \rangle_{2\pi} + 2ab\varepsilon^2\langle \sin(t)\cos(t) \rangle_{2\pi}$<br>$= (b-a)\varepsilon^2 > 0$<br>$\{\mathcal{T}_2 = \langle -2xf \rangle_{2\pi} = \langle -2x(-(b-a)\dot{x} - abx) \rangle_{2\pi} = ab\varepsilon^2 > 0\}$ | Unstable (Criterion I) |
| 5 | $\ddot{x} - (b-a)\dot{x} - abx = 0$ | $\lambda_1 = -a$<br>$\lambda_2 = b$<br>($0 \leq b < a$) | $\mathcal{T}_1 = \langle -2\dot{x}f \rangle_{2\pi} = \langle -2\dot{x}(-(b-a)\dot{x} - abx) \rangle_{2\pi}$<br>$= -2(b-a)\langle \dot{x}^2 \rangle_{2\pi} + 2ab\langle \dot{x}x \rangle_{2\pi}$<br>$= 2(b-a)\varepsilon^2\langle \sin^2(t) \rangle_{2\pi} + 2ab\varepsilon^2\langle \sin(t)\cos(t) \rangle_{2\pi}$<br>$= (b-a)\varepsilon^2 < 0$<br>$\mathcal{T}_2 = \langle -2xf \rangle_{2\pi} = \langle -2x(-(b-a)\dot{x} - abx) \rangle_{2\pi}$<br>$= 0 + 2ab\varepsilon^2\langle \cos^2(t) \rangle_{2\pi} = ab\varepsilon^2 > 0$ | Unstable (Criterion II) |
| 6 | $\ddot{x} = 0$ | $\lambda_1 = \lambda_2 = 0$ | $\mathcal{T}_1 = \langle -2\dot{x}f \rangle_{2\pi} = \langle 0 \rangle_{2\pi} = 0$<br>$\mathcal{T}_2 = \langle -2xf \rangle_{2\pi} = \langle 0 \rangle_{2\pi} = 0$ | Unstable (Criterion V) |

**Table 5** Applying the proposed theorem to different possible stable linear $2^{ed}$ order systems

| No | System | Eigen values | Sign of $\mathcal{T}_1$ and $\mathcal{T}_2$ (For $x = \varepsilon\cos(t)$ and $\dot{x} = -\varepsilon\sin(t)$) | Origin Status based on the proposed theorem |
|---|---|---|---|---|
| 1 | $\ddot{x} + (a+b)\dot{x} + abx = 0$ | $\lambda_1 = -a$<br>$\lambda_2 = -b$ | $\mathcal{T}_1 = \langle -2\dot{x}f \rangle_{2\pi} = \langle -2\dot{x}((a+b)\dot{x} + abx) \rangle_{2\pi}$<br>$= -2(a+b)\langle \dot{x}^2 \rangle_{2\pi} - 2ab\langle \dot{x}x \rangle_{2\pi}$ | Asymptotically Stable |

| | | (0<a<b) | $= -2(a+b)\varepsilon^2\langle\sin^2(t)\rangle_{2\pi} - 2ab\varepsilon^2\langle\sin(t)\cos(t)\rangle_{2\pi}$ $= -(a+b)\varepsilon^2 < 0$ $\mathcal{T}_2 = \langle -2xf\rangle_{2\pi} = \langle -2x((a+b)\dot{x}+abx)\rangle_{2\pi}$ $= 0 - 2ab\varepsilon^2\langle\cos^2(t)\rangle_{2\pi}$ $= -ab\varepsilon^2 < 0$ | (Criterion III) |
|---|---|---|---|---|
| 2 | $\ddot{x} + 2a\dot{x} + a^2x = 0$ | $\lambda_1=\lambda_2= -a$ (0< a) | $\mathcal{T}_1 = \langle -2\dot{x}f\rangle_{2\pi} = \langle -2\dot{x}(2a\dot{x}+a^2x)\rangle_{2\pi}$ $= -4a\varepsilon^2\langle\sin^2(t)\rangle_{2\pi} = -2a\varepsilon^2 < 0$ $\mathcal{T}_2 = \langle -2xf\rangle_{2\pi} = \langle -2x(2a\dot{x}+a^2x)\rangle_{2\pi}$ $= 0 - 2a^2\varepsilon^2\langle\cos^2(t)\rangle_{2\pi} = -a^2\varepsilon^2 < 0$ | Asymptotically Stable (Criterion III) |
| 3 | $\ddot{x} + 2a\dot{x} + (a^2 + b^2)x = 0$ | $\lambda_1=-a+b\mathrm{j}$ $\lambda_2=-a-b\mathrm{j}$ (0<a, 0<b) | $\mathcal{T}_1 = \langle -2\dot{x}f\rangle_{2\pi} = \langle -2\dot{x}(2a\dot{x}+(a^2+b^2)x)\rangle_{2\pi}$ $= -4a\langle\dot{x}^2\rangle_{2\pi} - 2(a^2+b^2)\langle\dot{x}x\rangle_{2\pi}$ $= -4a\varepsilon^2\langle\sin^2(t)\rangle_{2\pi} - 2\varepsilon^2(a^2+b^2)\langle\sin(t)\cos(t)\rangle_{2\pi}$ $= -2a\varepsilon^2 < 0$ $\mathcal{T}_2 = \langle -2xf\rangle_{2\pi} = \langle -2x(2a\dot{x}+(a^2+b^2)x)\rangle_{2\pi}$ $= 0 - 2(a^2+b^2)\varepsilon^2\langle\cos^2(t)\rangle_{2\pi} = -(a^2+b^2)\varepsilon^2 < 0$ | Asymptotically Stable (Criterion III) |
| 4 | $\ddot{x} + bx = 0$ | $\lambda_1=b\mathrm{j}$ $\lambda_2=-b\mathrm{j}$ (0<b) | $\mathcal{T}_1 = \langle -2\dot{x}f\rangle_{2\pi} = \langle -2\dot{x}bx\rangle_{2\pi}$ $= -2b\varepsilon^2\langle\sin(t)\cos(t)\rangle_{2\pi} = 0$ $\mathcal{T}_2 = \langle -2xf\rangle_{2\pi} = \langle -2x(bx)\rangle_{2\pi}$ $=-2b\varepsilon^2\langle\cos^2(t)\rangle_{2\pi} = -b\varepsilon^2 < 0$ | Marginally Stable (Criterion IV) |

**Table 6** Type of singular points based on the proposed criteria

| Equilibrium status | Eigenvalues | Values of ($\mathcal{T}_1$, $\mathcal{T}_2$) | Type of singular point |
|---|---|---|---|
| **Asymptotically Stable** | $\lambda_1= -a$  $\lambda_2= -b$ (0<a<b) | $\mathcal{T}_1 = -(a+b)\varepsilon^2 < 0$ $\mathcal{T}_2 = -ab\varepsilon^2 < 0$ | Node |
| | $\lambda_1=\lambda_2= -a$ (0< a) | $\mathcal{T}_1 = -2a\varepsilon^2 < 0$ $\mathcal{T}_2 = -a^2\varepsilon^2 < 0$ | Degenerate Node |
| | $\lambda_1=-a+b\mathrm{j}$  $\lambda_2= -a-b\mathrm{j}$ (0<a, 0<b) | $\mathcal{T}_1 = -2a\varepsilon^2 < 0$ $\mathcal{T}_2 = -(a^2+b^2)\varepsilon^2 < 0$ | Focus (spiral) |
| **Marginally Stable** | $\lambda_1=b\mathrm{j}$  $\lambda_2=-b\mathrm{j}$ (0<b) | $\mathcal{T}_1 = 0$ $\mathcal{T}_2 = -b\varepsilon^2 < 0$ | Center |
| **Unstable** | $\lambda_1=a$  $\lambda_2=b$ (0<a<b) | $\mathcal{T}_1 = (a+b)\varepsilon^2 > 0$ $\{\mathcal{T}_2 = -ab\varepsilon^2 < 0\}$ | Node |
| | $\lambda_1=\lambda_2= a$ (0< a) | $\mathcal{T}_1 = 2a\varepsilon^2 > 0$ $\{\mathcal{T}_2 = -a^2\varepsilon^2 < 0\}$ | Degenerate Node |
| | $\lambda_1=a+b\mathrm{j}$  $\lambda_2=a-b\mathrm{j}$ (0<a, 0<b) | $\mathcal{T}_1 = 2a\varepsilon^2 > 0$ $\{\mathcal{T}_2 = -(a^2+b^2)\varepsilon^2 < 0\}$ | Focus (spiral) |
| | $\lambda_1= -a$  $\lambda_2= b$ (0≤a<b) | $\mathcal{T}_1 = (b-a)\varepsilon^2 > 0$ $\{\mathcal{T}_2 = ab\varepsilon^2 > 0\}$ | Saddle |
| | $\lambda_1= -a$  $\lambda_2= b$ (0≤b<a) | $\mathcal{T}_1 = (b-a)\varepsilon^2 < 0$ $\mathcal{T}_2 = ab\varepsilon^2 > 0$ | Saddle |

| | $\lambda_1 = \lambda_2 = 0$ | $\mathcal{T}_1 = 0$ $\mathcal{T}_2 = 0$ | Saddle (Uniform motion) |
|---|---|---|---|

*6.2. Applying the stability theorem to $2^{ed}$-order nonlinear systems*

**Van der Pol oscillator.** Until now, the effectiveness of the proposed theorem was demonstrated for linear differential equations. Next, an important question may arise about the effectiveness of the proposed theorem for nonlinear dynamical systems. Consequently, some well-known nonlinear systems in which the equilibrium state is the origin of the phase space are investigated. The Van der Pol oscillator is a well-known nonlinear system presented by the Dutch physicist Balthasar Van der Pol [16, 17]. The Van der Pol equation given in Table 7 has an unstable equilibrium point at the origin according to the Poincaré theorem [2 ,4] which expresses a system will experience uniform periodic motion, if the enclosed equilibrium point is unstable. Now, let's study the instability of the equilibrium point in the Van der Pol equation via the proposed theorem. As can be clearly seen in the first raw of Table 7, the equilibrium point of the Van der Pol oscillator is unstable as $\mathcal{T}_1 = \mu(\varepsilon^2 - \frac{\varepsilon^4}{4}) > 0$ for small $\varepsilon$ (i.e. $\varepsilon < 2$) based on the presented theorem. In addition to that, it is interesting to note that the functional $\mathcal{T}_1$ can also predict the existence of limit cycle in the Van der Pol oscillator response corresponding to the oscillation amplitude of $\varepsilon = 2$ at which $\mathcal{T}_1 = \mu(\varepsilon^2 - \frac{\varepsilon^4}{4}) = 0$. Although the exact limit cycle of the Van der Pol oscillator is, different from the considered orbit (i.e. $x = 2\cos(t)$ and $\dot{x} = -2\sin(t)$) in the proposed theorem it is worth noting that the motion amplitudes are the same for both cases. Consequently, the first functional $\mathcal{T}_1$ returns a zero value corresponding to the orbit of the same amplitude as the realistic limit cycle of the system. Accordingly, the necessary condition for limit cycle existence would be $\mathcal{T}_1 = 0$ based on the presented theorem.

**Duffing oscillator.** More evidence about the effectiveness of the proposed theorem for analysis of nonlinear systems can be found in the rest of cases considered in Table 7. The second example pertains to the Duffing oscillator with nonlinear damping [23] which has been proved to be stable around the origin in [4] using the standard Lyapunov theorem. Clearly, the proposed theorem also affirms the stability of the origin (as $\mathcal{T}_1$ and $\mathcal{T}_2$ are both negative) without the requirement to search for a Lyapunov function.

**Pendulum system.** Finally, the third and fourth case studies given in Table 7 pertain to the well-known nonlinear pendulum system containing infinite number of equilibrium points such as $\mathbf{x_{e1}}$=(0,0), $\mathbf{x_{e2}}$=(π,0)), etc. As seen in Table 7, the presented theorem affirms $\mathbf{x_{e1}}$ is asymptotically stable (as $\mathcal{T}_1 < 0$ and $\mathcal{T}_2 < 0$) while $\mathbf{x_{e2}}$ is unstable (as as $\mathcal{T}_1 < 0$ and $\mathcal{T}_2 > 0$). As mentioned earlier, a prominent characteristic of the presented theorem was to determine the type of singular points. Then, let's determine the type of singular points in the pendulum system accordingly. As can be observed in the third raw of Table 7, the amounts of functional $\mathcal{T}_1$ and $\mathcal{T}_2$ were calculated as $-\varepsilon^2$ and $-\varepsilon^2$ respectively. Subsequently, using Table 6, one can easily find two real positive values for parameters *a* and *b* (i.e. $\frac{1}{2}$ and $\frac{\sqrt{3}}{2}$ respectivly) merely based on the set of equations $\mathcal{T}_1 = -2a\varepsilon^2 = -\varepsilon^2$ and $\mathcal{T}_2 = -(a^2 + b^2)\varepsilon^2 = -\varepsilon^2$ which implies that the first equilibrium point ($\mathbf{x_{e1}}$) is a stable focus (see the third row of Table 6). However, the values of functional $\mathcal{T}_1$ and $\mathcal{T}_2$ were found to be $-\varepsilon^2$ and $+\varepsilon^2$ respectively for the second equilibrium point ($\mathbf{x_{e2}}$). Thus, using Table 6, *a*=2.736 and *b*=0.736 considering the eighth row based on the set of equations $\mathcal{T}_1 = (b-a)\varepsilon^2 = -\varepsilon^2$ and $\mathcal{T}_2 = ab\varepsilon^2 = +\varepsilon^2$ which can lead one into believing the second equilibrium point ($\mathbf{x_{e2}}$) is a saddle point.

**Table 7** Applying the presented theorem to different $2^{ed}$ order nonlinear systems

| Case No. | System | Sign of $\mathcal{T}_1$ and $\mathcal{T}_2$ (For $x = \varepsilon\cos(t)$ and $\dot{x} = -\varepsilon\sin(t)$) | Status of origin |
|---|---|---|---|
| 1 | $\ddot{x} - \mu(1-x^2)\dot{x} + x = 0$ (where μ>0) | $\mathcal{T}_1 = \langle -2\dot{x}f\rangle_{2\pi} = \langle -2\dot{x}(-\mu(1-x^2)\dot{x} + x)\rangle_{2\pi}$ $= 2\mu\langle \dot{x}^2\rangle_{2\pi} - 2\mu\langle \dot{x}^2 x^2\rangle_{2\pi} - 2\langle x\dot{x}\rangle_{2\pi}$ $= 2\mu\varepsilon^2\langle \sin^2(t)\rangle_{2\pi} - 2\mu\varepsilon^4\langle \sin^2(t)\cos^2(t)\rangle_{2\pi} - 0$ $= \mu(\varepsilon^2 - \frac{\varepsilon^4}{4}) > 0$ (ε is a small positive value) $\{\mathcal{T}_2 = \langle -2xf\rangle_{2\pi} = \langle -2x(-\mu(1-x^2)\dot{x} + x)\rangle_{2\pi} = -\varepsilon^2 < 0\}$ | Unstable (Criterion I) |
| 2 | $\ddot{x} + |\dot{x}|\dot{x} + 3x + x^3 = 0$ | $\mathcal{T}_1 = \langle -2\dot{x}f\rangle_{2\pi} = \langle -2\dot{x}(|\dot{x}|\dot{x} + 3x + x^3)\rangle_{2\pi}$ $= -2\langle |\dot{x}|\dot{x}^2\rangle_{2\pi} - 6\langle \dot{x}x\rangle_{2\pi} - 2\langle \dot{x}x^3\rangle_{2\pi}$ $= -2\varepsilon^3\langle |\sin(t)|\sin^2(t)\rangle_{2\pi} - 0 - 2\varepsilon^4\langle \sin(t)\cos^3(t)\rangle_{2\pi}$ $= -0.85\varepsilon^3 < 0$ (ε is a small positive value) $\mathcal{T}_2 = \langle -2xf\rangle_{2\pi} = \langle -2x(|\dot{x}|\dot{x} + 3x + x^3)\rangle_{2\pi}$ $=+2\varepsilon^3\langle |\sin(t)|sin(t)\cos(t)\rangle_{2\pi}-6\varepsilon^2\langle \cos^2(t)\rangle_{2\pi}$ $-2\varepsilon^4\langle \cos^4(t)\rangle_{2\pi} = 0 - 3\varepsilon^2 - 0.75\varepsilon^4 < 0$ | Stable (Criterion III) |
| 3.1 | $\ddot{x} + \dot{x} + \sin(x) = 0$ ($x_{e1}$=(0,0)) | $\mathcal{T}_1 = \langle -2\dot{x}f\rangle_{2\pi} = \langle -2\dot{x}(\dot{x} + \sin(x))\rangle_{2\pi}$ $= -2\langle \dot{x}^2\rangle_{2\pi} - 2\langle \dot{x}\sin(x)\rangle_{2\pi}$ $= -2\varepsilon^2\langle \sin^2(t)\rangle_{2\pi} + 2\varepsilon\langle \sin(t)\sin(\varepsilon\cos(t))\rangle_{2\pi} = -\varepsilon^2 < 0$ (ε is a small positive value) $\mathcal{T}_2 = \langle -2xf\rangle_{2\pi} = \langle -2x(\dot{x} + \sin(x))\rangle_{2\pi}$ $=-2\varepsilon\langle \cos(t)\sin(\varepsilon\cos(t))\rangle_{2\pi} = -\varepsilon^2 < 0$ (ε is a small positive value) | Stable (Criterion III) |
| 3.2 | $\ddot{x} + \dot{x} + \sin(x) = 0$ ($x_{e2}$=(π,0)) | First, shifting the equilibrium state to the origin (using z=x- π) yields $\ddot{z} + \dot{z} - \sin(z) = 0$ $\mathcal{T}_1 = \langle -2\dot{z}f\rangle_{2\pi} = \langle -2\dot{z}(\dot{z} - \sin(z))\rangle_{2\pi}$ $= -2\langle \dot{z}^2\rangle_{2\pi} - 2\langle \dot{z}\sin(z)\rangle_{2\pi}$ $= -2\varepsilon^2\langle \sin^2(t)\rangle_{2\pi} + 2\varepsilon\langle \sin(t)\sin(\varepsilon\cos(t))\rangle_{2\pi} = -\varepsilon^2 < 0$ (ε is a small positive value) $\mathcal{T}_2 = \langle -2zf\rangle_{2\pi} = \langle -2z(\dot{z} - \sin(z))\rangle_{2\pi}$ $=0 + 2\varepsilon\langle \cos(t)\sin(\varepsilon\cos(t))\rangle_{2\pi} = \varepsilon^2 > 0$ (ε is a small positive value) | Unstable (Criterion II) |

*6.3. Linearization of a $2^{ed}$-order nonlinear system via averaging*

The last example of the paper is devoted to linearization of a special nonlinear system around the origin based on the proposed averaging-based approach. Consider a nonlinear system as

$$\dot{x}_1 = x_2 \tag{41}$$

$$\dot{x}_2 = x_2^2 - x_1 + x_1^3 \tag{42}$$

Clearly, the origin of the phase plane is an equilibrium state of the nonlinear system. First, let's linearize the system around the origin using Jacobian matrix as

$$\mathbf{A} = \begin{bmatrix} \frac{\partial f_1}{\partial x_1} & \frac{\partial f_1}{\partial x_2} \\ \frac{\partial f_2}{\partial x_1} & \frac{\partial f_2}{\partial x_2} \end{bmatrix} = \begin{bmatrix} 0 & 1 \\ -1 + 3x_1^2 & 2x_2 \end{bmatrix} = \begin{bmatrix} 0 & 1 \\ -1 & 0 \end{bmatrix}_{x_e=(0,0)} \tag{43}$$

Accordingly, the eigenvalues of **A** are found to be +1j and -1j. Hence, the linearized system is marginally stable and thus, one cannot determine the stability or instability of the original nonlinear system. Now, let's check this latest issue via the proposed averaging-based linearization technique. Thus, using Eq. (33) and applying a very small excitation orbit enclosing the origin, one can write the state matrix as

$$\mathbf{A} = \operatorname*{Lim}_{\varepsilon \to 0^+} \frac{2}{\varepsilon^2} \begin{bmatrix} \langle x_1 f_1 \rangle_{2\pi} & \langle x_2 f_1 \rangle_{2\pi} \\ \langle x_1 f_2 \rangle_{2\pi} & \langle x_2 f_2 \rangle_{2\pi} \end{bmatrix} = \operatorname*{Lim}_{\varepsilon \to 0^+} \frac{2}{\varepsilon^2} \begin{bmatrix} \langle x_1 x_2 \rangle_2 & \langle x_2^2 \rangle_{2\pi} \\ \langle x_1(x_2^2 - x_1 + x_1^3) \rangle_{2\pi} & \langle x_2(x_2^2 - x_1 + x_1^3) \rangle_{2\pi} \end{bmatrix}$$

$$= \operatorname*{Lim}_{\varepsilon \to 0^+} \begin{bmatrix} 0 & 1 \\ -1 + \frac{3}{4}\varepsilon^2 & 0 \end{bmatrix} = \begin{bmatrix} 0 & 1 \\ -1 & 0 \end{bmatrix} \tag{44}$$

As seen, the linearized system by the proposed averaging-based technique is identical to that of the Jacobian matrix once $\varepsilon \to 0^+$ meaning that the linearized system is marginally stable. However, it is interesting to note that for $0 < \varepsilon < \frac{2}{\sqrt{3}}$ the eigenvalues of **A** given in Eq. (44) remain on the imaginary axis meaning that the original nonlinear system will be marginally stable around the origin. Finally, Fig. 2 shows the phase portrait of the original nonlinear system (i.e. Eqs. (41) and (42)), which affirms the marginal stability of the nonlinear system at the vicinity of the origin through which the results of the proposed stability theorem is validated. Indeed, the mentioned finding is a significant benefit of the proposed linearization technique compared to the Jacobian method.

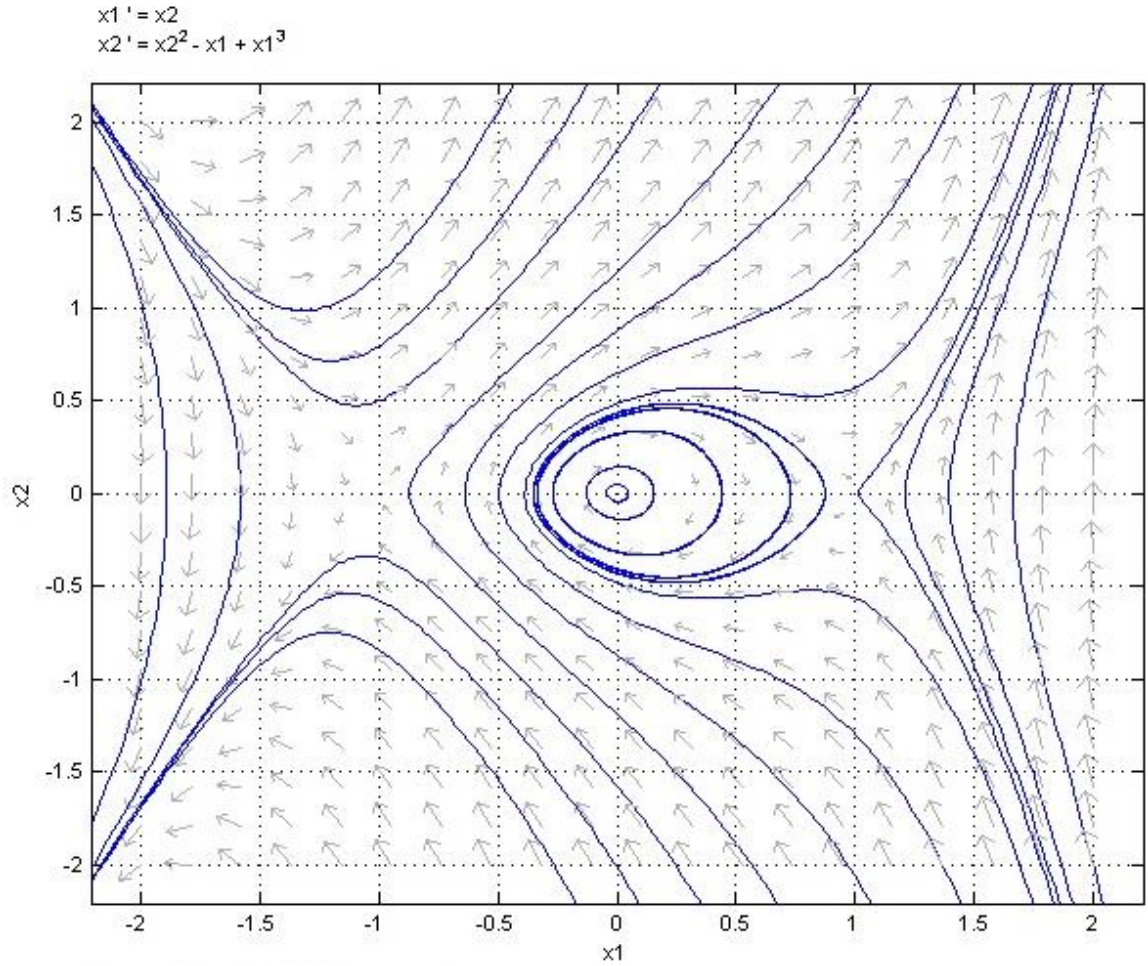

**Fig. 2.** Phase portrait for Eqs. (41) and (42)

## 7. Conclusions

In this work, a new approach for determination of equilibrium status at singular points of second order systems was introduced. It was found that the presented idea is applicable to both linear and nonlinear dynamical systems as well. Besides, the proposed theorem could be used to study both instability and stability problems. In comparison with the direct Lyapunov method that relied on finding a specific Lyapunov function in order to prove stability/instability (which was often hard to obtain), the presented theorem proposed two distinct functionals as well as some straightforward criteria that could be applied to all linear and nonlinear problems. Thus, the challenge of selecting a suitable Lyapunov function for complex dynamic problems was resolved using the proposed idea. Besides, the type of singular points in nonlinear dynamical systems such as center, saddle, node and focus could be predicted. Then, the presented theorem could be extended for linearization of nonlinear systems around the origin based on the averaging principle as a counterpart to the Lyapunov indirect method. Next, the superiority of the proposed integration-based linearization method over the indirect Lyapunov method was demonstrated. Predictably, the presented theorem can be extensively employed in different fields of engineering and science such as nonlinear dynamics, nonlinear vibration, nonlinear dynamical systems, nonlinear control theory, oscillators, engines, energy absorbers, energy harvesters etc.

## Acknowledgements

The author would like to acknowledge the Shiraz University of Technology for providing research facilities and supports.